\documentclass[twocolumn,amsmath,amssymb,prl,10pt,nofootinbib,superscriptaddress]{revtex4}
\usepackage{ graphicx, float,amsmath, amsmath, amssymb}
\usepackage[export]{adjustbox}

\def\be{\begin{equation}}
\def\ee{\end{equation}}
\def\bea{\begin{eqnarray}}
\def\eea{\end{eqnarray}}
\def\bse{\begin{subequations}}
\def\ese{\end{subequations}}

\usepackage[breaklinks, colorlinks, citecolor=blue]{hyperref}
\usepackage[labelsep=period]{caption}
\usepackage[normalem]{ulem}
\usepackage{soul}
\usepackage{minitoc}

\usepackage{bm}

\begin{document}

\title{Quasinormal modes of black holes in a toy-model for cumulative quantum gravity}

\author{Aur\'elien Barrau}%
\affiliation{%
Laboratoire de Physique Subatomique et de Cosmologie, Universit\'e Grenoble-Alpes, CNRS/IN2P3\\
53, avenue des Martyrs, 38026 Grenoble cedex, France
}

\author{Killian Martineau}%
\affiliation{%
Laboratoire de Physique Subatomique et de Cosmologie, Universit\'e Grenoble-Alpes, CNRS/IN2P3\\
53, avenue des Martyrs, 38026 Grenoble cedex, France
}

\author{Jeremy Martinon}%
\affiliation{%
Laboratoire de Physique Subatomique et de Cosmologie, Universit\'e Grenoble-Alpes, CNRS/IN2P3\\
53, avenue des Martyrs, 38026 Grenoble cedex, France
}


\author{Flora Moulin}%
\affiliation{%
Laboratoire de Physique Subatomique et de Cosmologie, Universit\'e Grenoble-Alpes, CNRS/IN2P3\\
53, avenue des Martyrs, 38026 Grenoble cedex, France
}


\date{\today}
\begin{abstract} 
The idea that quantum gravity effects might ``leak" outside the horizon of a black hole has recently been intensively considered. In this study, we calculate the quasinormal modes as a function of the location and amplitude of a generic metric perturbation distorting to the Schwarzschild spacetime. We conclude on the possible observability of quantum metric corrections by current and future gravitational wave experiments.
\end{abstract}
\maketitle

\section{Introduction}

Naively, quantum gravity is expected to show up at very small physical scales, around the Planck length (see \cite{Barrau:2017tcd} for a recent review of the phenomenology of quantum gravity). This is indeed where predictions become precise and might lead to a clear discrimination between models. In the black hole (BH) sector, it has therefore been widely believed that quantum gravity effects are confined to the vicinity of the central singularity. This is clearly the most conservative and natural hypothesis. In such a case, quantum modifications to the spacetime structure are screened by the event horizon and the external observer is not expected to notice any measurable effect, at least for macroscopic black holes. In this article, we focus on a different perspective, namely the possibility that quantum corrections to the metric ``leak" outside the horizon, even for stellar or supermassive BHs. This is obviously motivated by phenomenological reasons. There are, however, quite good physical motivations to consider quantum gravity effects well beyond the vicinity of the singularity. Studying their impact, in a very simple model, on the ringdown phase of BHs is the purpose of this study.\\

It has recently been argued in \cite{Giddings:2016btb} and \cite{Giddings:2019jwy} that the observation of black holes with the Event Horizon Telescope might reveal quantum gravity effects. Consistency between general relativity (GR) and quantum mechanics (QM) might require quantum effects at very large scale. Interestingly, the authors suggest that the time dependence of the shape and size of the shadow that a black hole casts on its surrounding emission might be seen around the BH at the center of the M87 galaxy (which has recently been effectively observed \cite{Akiyama:2019cqa,Akiyama:2019eap}). On the extreme other side, in the firewall proposal, the usual geometry might break down a Planck length away from the horizon \cite{Almheiri:2012rt,Almheiri:2013hfa}. Many other possibilities with strong metric modifications outside the horizon (or what replaces it) have been considered: gravastars \cite{Mazur:2004fk}, fuzzballs where string theory configurations replace the smooth manifold outside the horizon \cite{Mathur:2008nj}, or massive remnants \cite{Giddings:1992hh}. The study of maximally entangled states of black holes has even shed a new light on the possibility of more drastic geometric effects far away from the horizon \cite{Maldacena:2013xja}. To give a final example, bouncing black holes -- with quite different time-scales -- are also intensively considered \cite{Barcelo:2017lnx,Haggard:2014rza}.\\

In this article, we focus on a different approach which is based on heuristic considerations \cite{Haggard:2016ibp}. This is to be considered as a toy-model or a kind of ``prototype" of what could be expected in optimistic quantum gravity scenarii. Our aim is to calculate the displacement of quasinormal modes and quantify the amplitude of the metric modification that would be required for an experimental detection. This might be used beyond this specific model. Focusing only on non-rotating BHs we do not search for accurate results, that would be meaningless at this stage, but just try to estimate the orders of magnitude for future studies. In the next section we briefly explain the method used to evaluate the frequency and amplitude of the ringing modes of BHs. Then, we explain the model used and explicitly show our results. 

\section{Quasinormal modes}

Quasinormal modes (QNMs) are the decaying modes of black holes. As BHs are vacuum solutions of the Einstein field equation, QNMs can be regarded as the intrinsic vibrationnal and damping properties of spacetime itself. After a BH has been perturbed, three phases can be distinguished: the transient event, the quasinormal mode ringdown, and the damped tail. \\

The ringdown phase of a BH does not lead to precisely ``normal" modes because the system looses energy through gravitational waves. The wave equation for the metric perturbation is unusual because of its boundary conditions: the wave should be purely outgoing at infinity and purely ingoing at the BH horizon. The radial part of the oscillation can be written (see \cite{Chirenti:2017mwe} for an intuitive introductory review) as $\phi \propto e^{-i\omega t} = e^{-i(\omega_R + i\omega_I)t}$ where the complex pulsation $\omega$ decomposes in a real part $\omega_R$ and an imaginary part $\omega_I$, which is the inverse timescale of the damping. The process is stable only when $\omega_I<0$. Technically, the calculation of QNMs is quite reminiscent of the one of greybody factors (see, {\it e.g.}, \cite{Moulin:2018uap} for a recent derivation with a quantum-gravity modified metric) which describes the scattering of quantum fields in a BH background.\\

The perturbations of the Schwarzschild metric are of two different types. One is called ``axial", it gives small values to the metric coefficients that were zero, inducing a frame dragging and rotation of the black hole. The other is called ``polar" and gives small increments to the already non-zero metric coefficients. They are governed by two different equations. 
Perturbations with the axial parity are given by the Regge-Wheeler equation with the potential
\begin{equation}
V^{\textrm{RG}}_{\ell}(r) = \left(1-\frac{2M}{r}\right)\left[\frac{\ell(\ell+1)}{r^2} - \frac{6M}{r^3}\right]\,,
\label{eq:RG}
\end{equation}
while perturbations with the polar parity are given by the Zerilli equation with potential
\begin{eqnarray}
& &V^{\textrm{Z}}_{\ell}(r) = \frac{2}{r^3}\left(1-\frac{2m}{r}\right)\times \nonumber\\
&\times&\frac{9M^3 + 3a^2Mr^2 + a^2(1+a)r^3 + 9M^2ar}{(3M+ar)^2}\,,
\label{eq:Ze}
\end{eqnarray}
where $a = \ell(\ell+1)/2 - 1$. For gravitational perturbations, one needs $\ell\ge2$. Importantly, those equations have the same spectrum of quasinormal modes. This isospectrality property \cite{Chandrasekhar:1985kt} is not always true in modified gravity (those considerations are well beyond the scope of this article and will be studied in another paper \cite{Flora}). Quasinomal modes are characterized by their overtone number $n$ and their multipole number $\ell$. For example, the fundamental quadrupolar mode ($n=0$ and $\ell=2$) for a Schwarzschild BH is given by $M\omega \approx 0.374 - 0.0890i$. \\

The calculation of quasinormal modes is nearly an art in itself (see \cite{Kokkotas:1999bd,Nollert:1999ji} for historical reviews and \cite{Berti:2004um,Dorband:2006gg} for an example of more recent results based on numerical approaches). In this study, we use a WKB approach described in \cite{Konoplya:2003ii} for D-dimensional BHs. The WKB method for QNMs was first introduced in \cite{Schutz:1985zz,Iyer:1986np,Iyer:1986nq,Kokkotas:1988fm} and has then be widely developed. The WKB formalism is very useful to obtain good approximations without having to rely on heavy numerical techniques. The higher the multipole number and the lower the overtone, the better the accuracy. We restrict ourselves to $n<l$ as the approximations otherwise break down. Details on the validity of the WKB approximation can be found in \cite{Schutz:1985zz} but, in any case, it requires the multipole number to be smaller than (or equal to, if the accuracy requirement is relaxed) the overtone number, otherwise the basic condition $|k'|\ll k^2$ (where $k^2$ is the potential of the considered effective Schr\"odinger equation) does not hold.\\

In order to have a good numerical accuracy, we have used the 6th order WKB method developed by Konoplya. It is presented in details in \cite{Konoplya:2003ii} (see also \cite{Konoplya:2019hlu}). This allows one to recast the potential appearing in the effective Schr\"odinger equation $(\frac{d^2\Psi}{dx^2}=k(x)\Psi(x))$ felt by gravitational perturbations in the form

\begin{equation}\label{2}
\frac{i k_{0}}{\sqrt{2 k_{0}''}}
-\Lambda_{2}-\Lambda_{3} -\Lambda_{4} -\Lambda_{5} -\Lambda_{6} =n+\frac{1}{2},
\end{equation}
where the terms $\Lambda_{i}$ are complicated -- but known -- expressions given in \cite{Konoplya:2003ii} whereas $k_{0}$ stands for the maximum of the potential and the derivative is to be understood with respect to the tortoise coordinate $r_*$ (defined by $dr_*=dr/f$ where $f$ is the metric function). 

\section{The model and its consequences}

We now focus on the toy model developed in \cite{Haggard:2016ibp}. The idea is very simple. The curvature scale is of the order of $l_R \sim {\cal R}^{-1/2}$, where the the Kretschmann scalar is ${\cal R}^2:=R_{\mu\nu\rho\lambda}R^{\mu\nu\rho\lambda}$. If one estimates the intensity of quantum gravitational effects through the ratio of Planck length over the curvature scale, the result is vanishingly small for stellar or supermassive BHs. This vision however disregards possible cumulative effects (also considered in \cite{Rovelli:2014cta,Barrau:2014hda,Barrau:2014yka,Barrau:2018kyv}). Dimensional arguments lead to the conclusion that the ``quantumness" of spacetime, integrated over a proper time $\tau$, might be given by $q=l_P  \ {\cal R} \ \tau$. As the proper time is related to the Schwarzschild time by
\begin{equation}
\tau=\sqrt{1-\frac{2M}{r}}\ t,
\end{equation}
one is led to 
\begin{equation}
q(r)  =  \frac{M}{r^3} \ \left(1-\frac{2M}{r}\right)^\frac{1}{2} t.
\end{equation}
Throughout all this study, we use Planck units. 
The maximum of this function is reached for $r=2M\left(1+\frac{1}{6}\right)$ and this is therefore where quantum gravity effects could be expected to be intense. \\

\begin{figure} 
\centering
    \includegraphics[width=.9\linewidth]{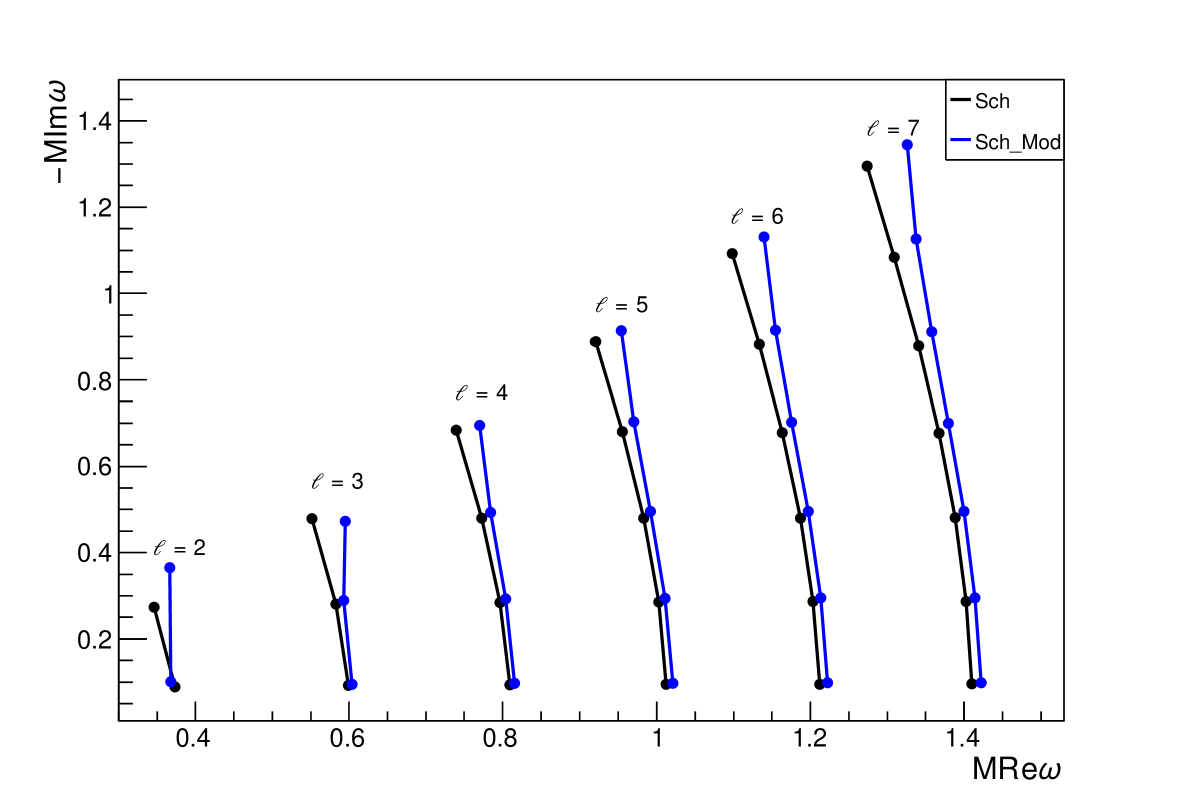}  
 \caption{Quasinormal mode complex frequencies for different multipolar orders (from $\ell=2$ to $\ell=7$ from the left to the right) and for different overtone numbers $n$ (increasing from the  lower points to the upper points). The black dots on the left correspond to the usual Schwarzschild case and the bleue dots on the right correspond to the modified metric with $\mu/M=2.3$ and $\sigma/M=1.5$} 
      \label{Fig0}
\end{figure}

The arguments previously given are obviously purely heuristic and should be considered as a rough indication of what might happen when time-integrated quantum corrections are optimistically considered. To remain quite generic, we parametrize a possible metric modification outside the horizon by a simple Gaussian function:
\begin{equation}
ds^2=-f(r)dt^2+f^{-1}(r)dr^2-r^2d\Omega^2,
\end{equation}
with
\begin{equation}
f(r)=\left(1-\frac{2M}{r}\right)\left(1+Ae^{-\frac{(r-\mu)^2}{2\sigma^2}}\right)^2.
\end{equation}

This Gaussian truncation of the Schwarzschild metric is {\it not} justified by any serious theoretical arguments. It should be seen as an effective metric encoding possible cumulative quantum effects outside the horizon. In addition it has the advantage not to shift the event horizon position. By varying the parameters $A,\mu,\sigma$, one can explore different shapes and positions for the ``quantum bump". In the following, we shall quantify the displacement of the real and imaginary parts of the QNMs as a function of the parameters ($\mu$ and $\sigma$ are expressed in units of $M$).\\

The complex frequencies are displayed in Fig. \ref{Fig0}. The black dots correspond to the general relativistic case whereas the blue ones, on the right, correspond to the considered modified case with $\mu=(7/6)R_S$, $A=0.01$, and $\sigma/M=1.5$. \\

\begin{figure} 
\centering
    \includegraphics[width=.9\linewidth]{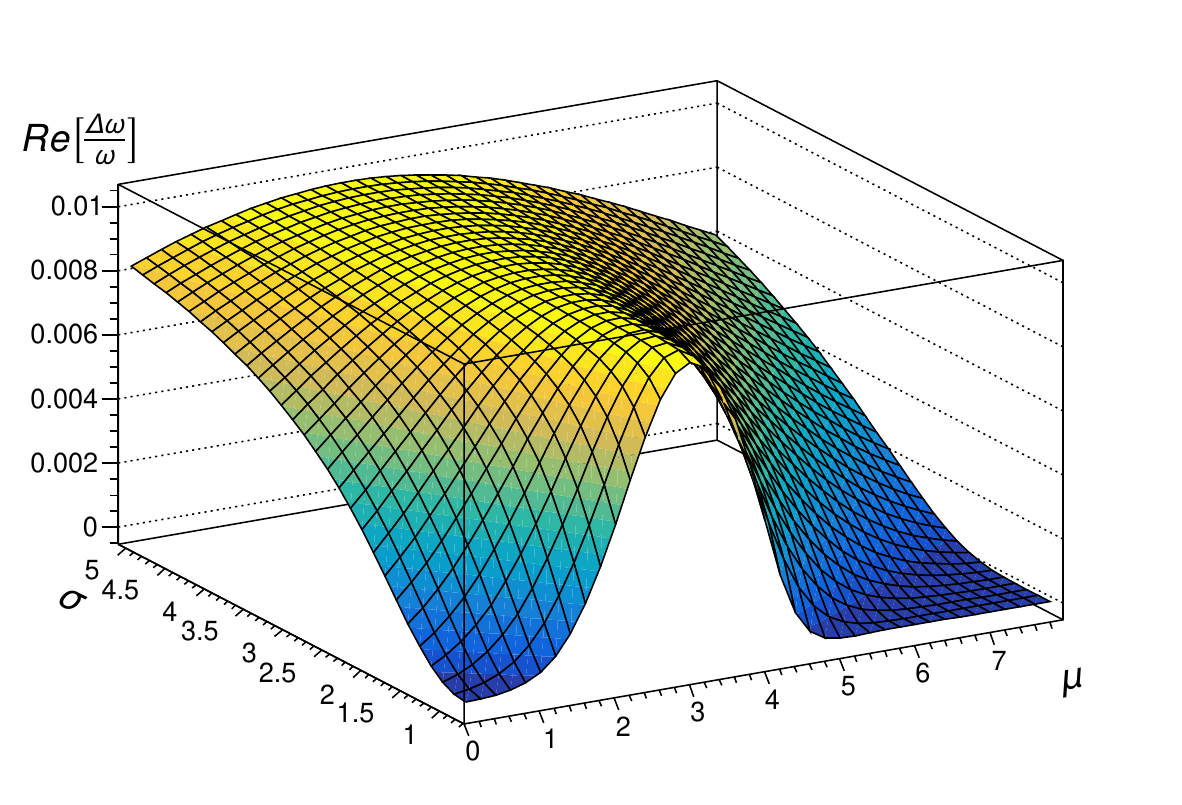}  
 \caption{Relative displacement of the real part of the quasinormal mode ($\ell=8,n=0$) as a function of $\mu$ and $\sigma$ for $A=0.01$.} 
      \label{Fig1}
\end{figure}

In Fig. \ref{Fig1}, the relative displacement of the real part of the quasinormal mode ($\ell=8,n=0$) is displayed as a function of $\mu$ and $\sigma$ for $A=0.01$. The trend does not radically depend on the specific mode chosen. We have therefore plotted here a quite high multipolar number as the WKB approximation is more reliable in this case. Interestingly -- but not that surprisingly -- it appears that the maximum displacement is obtained for $\mu \approx 3M$. In the limit of very large $l$, the value tends exactly to $3M$, which corresponds to the photon sphere and to the maximum of the potential. We have also considered in this figure a case where the maximum of the quantum correction is inside the horizon. Then, only the ``tail" of the Gaussian does affect the external spacetime. Even if the effect is smaller, it is still clearly non-vanishing. Interestingly the 2-dimensional surface is actually an ensemble of Gaussian functions whose width on the $\mu$ axis happens to be (non trivially) equal to the considered value of $\sigma$.\\

\begin{figure} 
\centering
    \includegraphics[width=.9\linewidth]{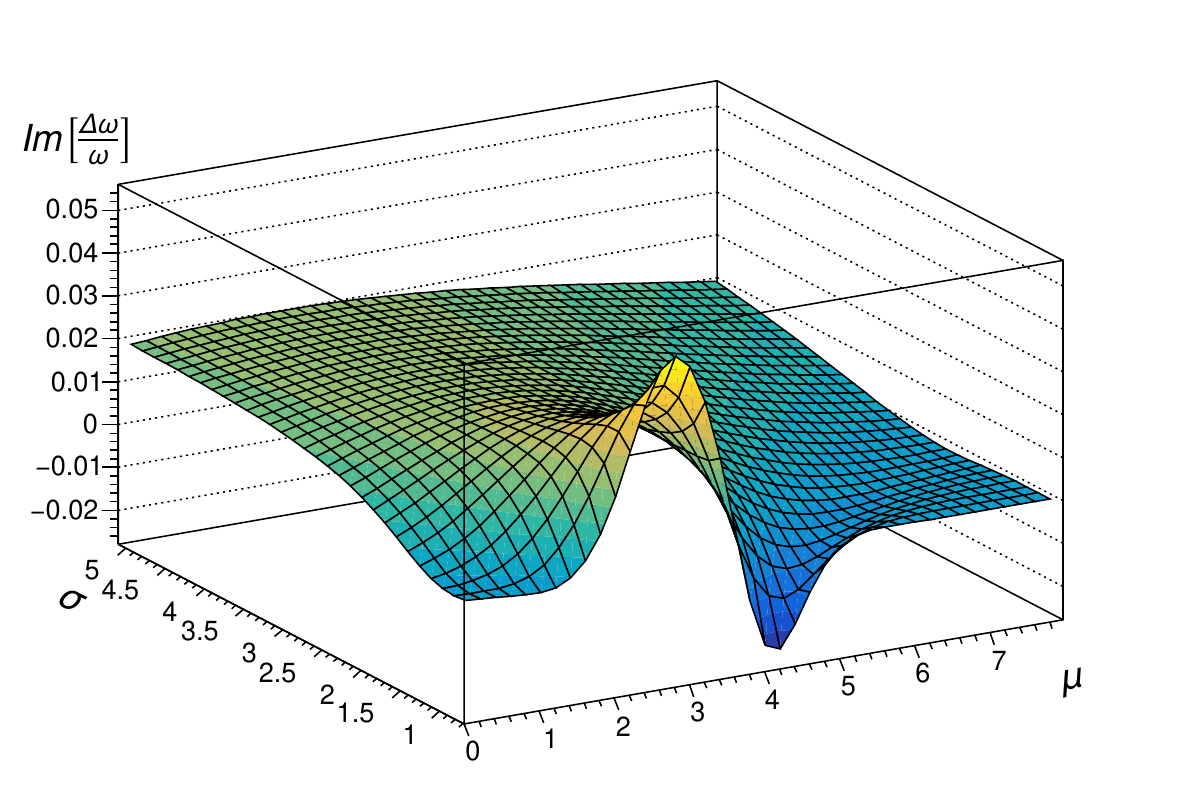}  
 \caption{Relative displacement of the imaginary part of the quasinormal mode ($\ell=8,n=0$) as a function of $\mu$ and $\sigma$ for $A=0.01$.} 
      \label{Fig2}
\end{figure}

In Fig. \ref{Fig2}, the relative displacement of the imaginary part of the quasinormal mode ($\ell=8,n=0$) is displayed as a function of $\mu$ and $\sigma$ for $A=0.01$. For quite low values of $\sigma$, the displacement can be either positive or negative for different values of $\mu$. This means that depending on its position the ``metric bump" can either increase on decrease the damping of gravitational waves.\\

\begin{figure} 
\centering
    \includegraphics[width=.9\linewidth]{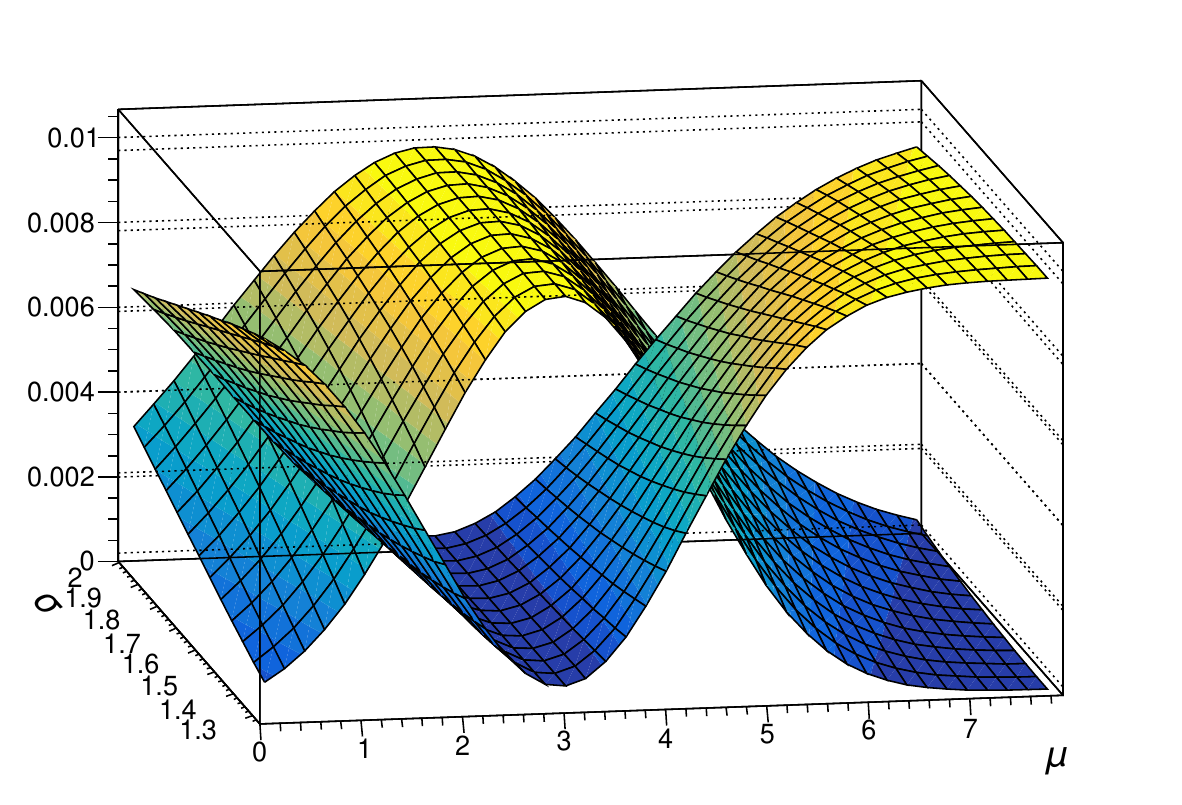}  
 \caption{Relative displacement of the real part of the quasinormal mode ($\ell=8,n=0$) as a function of $\mu$ and $\sigma$ for $A=0.01$ (upper curve at $\mu=3$) and $A=-0.01$.} 
      \label{Fig2.5}
\end{figure}

Finally, Fig. \ref{Fig2.5} shows the influence of the sign of the parameter $A$. The displacement is basically symmetrical.\\

For most of the considered parameter space, the displacement of the real part -- that is of the frequency -- is of the same order than the one of the imaginary part -- that is of the damping time. The easiest effect to measure is probably a frequency shift which happens to be always positive. Quite obviously, when the metric perturbation is very wide, its precise position looses any notable influence. 

\section{Observability}

Although gravitational waves have been ``detected" decades ago thanks to the Hulse-Taylor binary pulsar, the recent LIGO-Virgo detections (see \cite{Abbott:2016blz} for the seminal paper and \cite{LIGOScientific:2018mvr} for a first catalogue) have completely changed the game. Real astrophysical objects have spin and a modified Kerr solution should be considered, which is far beyond this prospective study. However, the global trends are expected to be the same and the orders of magnitude of the effects should be correct. Surprisingly, the very first event measured, GW150914, has already led to a detection of the fundamental quasinormal mode. It is not obvious to determine precisely the accuracy at which the characteristics of the QNMs are constrained by the current measurements. A relative accuracy of 50 \% is a conservative estimate. In the future, the Einstein Telescope (ET) should lead to a one order of magnitude better precision \cite{Sathyaprakash:2012jk}.\\

\begin{figure} 
\centering
    \includegraphics[width=.9\linewidth]{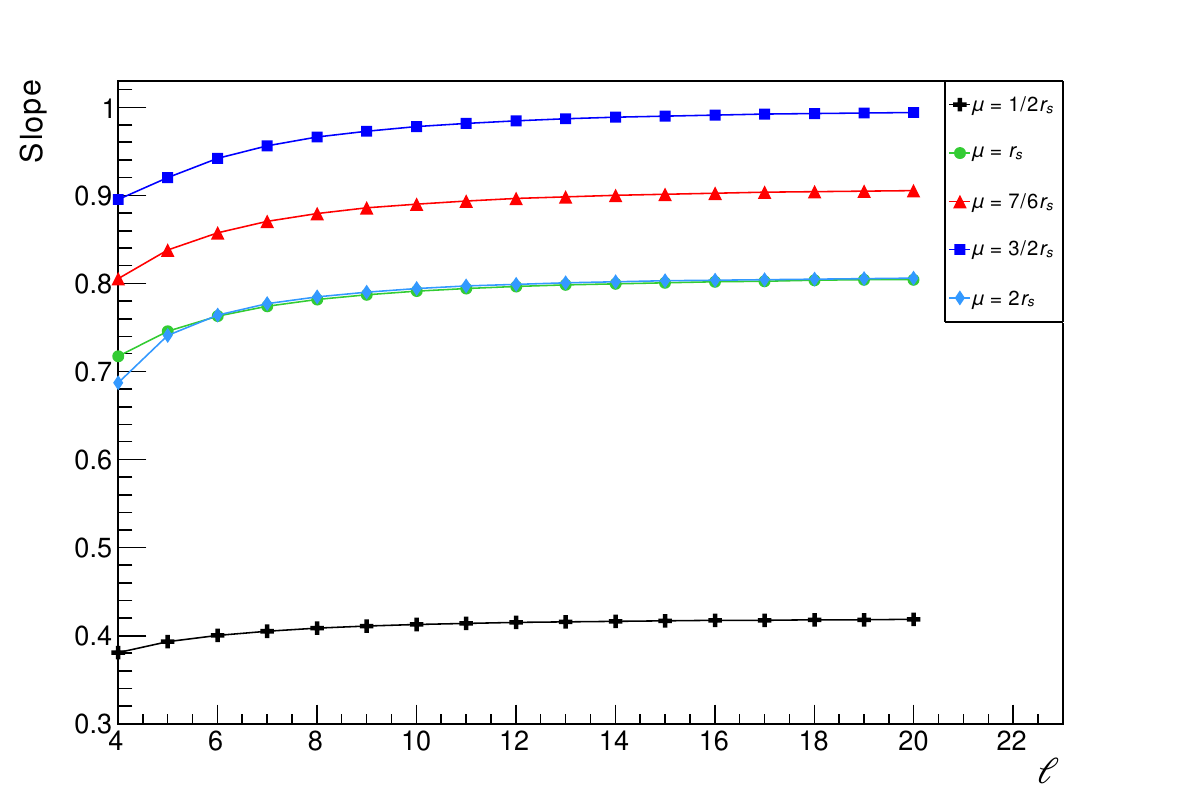}  
 \caption{Coefficient $x$ (such that $Re(\Delta\omega/\omega)=xA$) as a function of $\ell$ for $\sigma/M=1.5$. The different curves correspond to different values of $\mu$.} 
      \label{Fig3}
\end{figure}

\begin{figure} 
\centering
    \includegraphics[width=.9\linewidth]{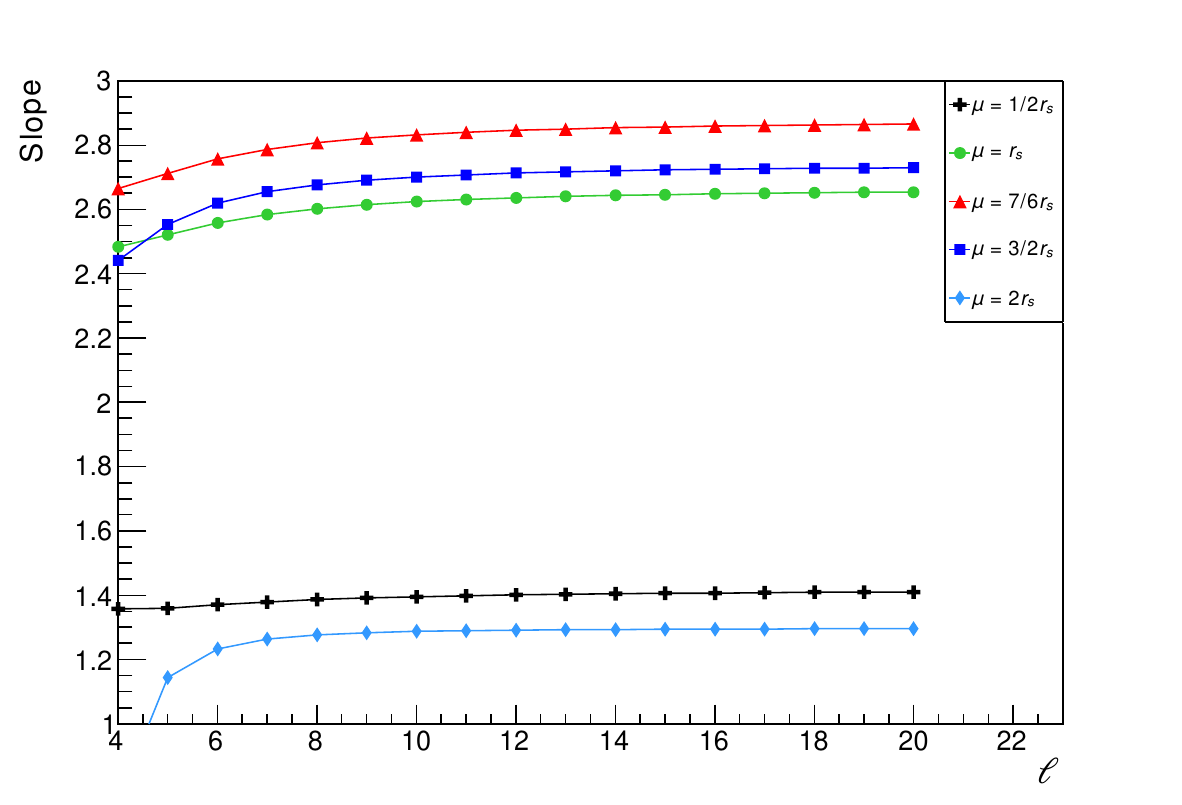}  
 \caption{Coefficient $y$ (such that $Im(\Delta\omega/\omega)=yA$) as a function of $\ell$ for $\sigma/M=1.5$. The different curves correspond to different values of $\mu$.} 
      \label{Fig4}
\end{figure}

\begin{figure} 
\centering
    \includegraphics[width=.9\linewidth]{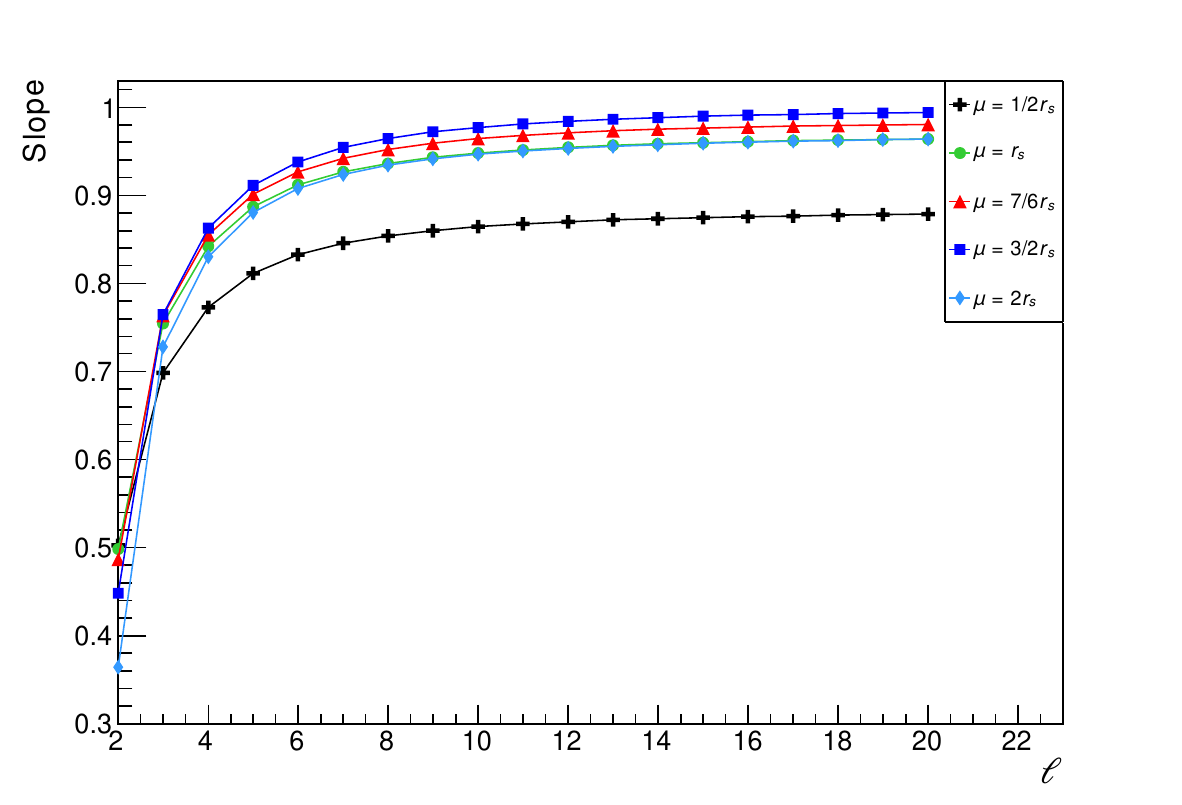}  
 \caption{Coefficient $x$ (such that $Re(\Delta\omega/\omega)=xA$) as a function of $\ell$ for $\sigma/M=4$. The different curves correspond to different values of $\mu$.} 
      \label{Fig5}
\end{figure}

\begin{figure} 
\centering
    \includegraphics[width=.9\linewidth]{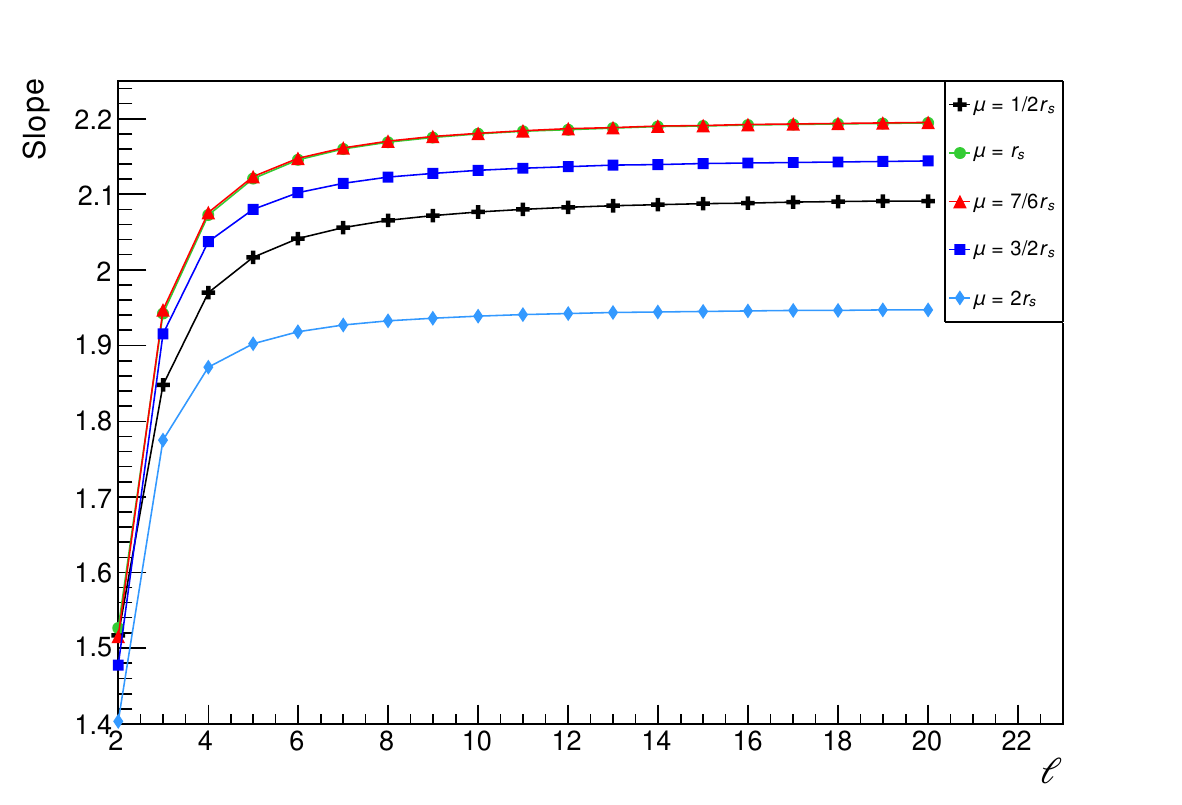}  
 \caption{Coefficient $y$ (such that $Im(\Delta\omega/\omega)=yA$) as a function of $\ell$ for $\sigma/M=4$. The different curves correspond to different values of $\mu$.} 
      \label{Fig6}
\end{figure}

The most important parameter for this study is obviously the constant $A$ which determines the amplitude of the correction. We have checked that the displacement of the QNMs complex frequency is linear as a function of $A$ over the interesting range. In Fig. \ref{Fig3}, we plot the slope of the real part of the QNM displacement versus $A$ ({\it i.e.} the $x$ parameter of $Re(\Delta\omega/\omega)=xA$) as a function of $\ell$, for $\sigma /M=1.5$. The different curves correspond to different values of the position $\mu$ of the quantum bump. In Fig. \ref{Fig4}, the very same thing is represented for the imaginary part of the QNM ({\it i.e.} the $y$ parameter of $Im(\Delta\omega/\omega)=yA$). In Fig. \ref{Fig5} and Fig. \ref{Fig6}, the value $\sigma /M=4$ is instead chosen. It should be pointed out that in some cases the lowest values of $\ell$ are not displayed as the WKB approximation breaks down and calculations could therefore be dubious.\\

Let us now get an order of magnitude of how those estimates relate to the toy model previously considered. As the $x$ and $y$ slopes are of order one, and as the relative displacement that could be measured is also of order one, this means that the $A$ parameter has to be of order unity so that the kind of quantum gravity effects studied here could be measured. If $A$ is assumed to be roughly comparable to the ``quantumness" $q$ introduced in the second section, one is led to the conclusion that $q$ should be of order one.  It is easy to check that 

\begin{equation}
q_{max}=\left(\frac{3}{7}\right)^3\sqrt{\frac{1}{7}}\frac{t}{M^2}.
\end{equation}

If one sets $t$ to be the age of the Universe, the mass value required so that the quantum gravity effects can be observed is of the order of $10^{-8}$ (or less) solar mass, that is roughly the mass of the Moon. Although far smaller than the mass of stellar black holes, this value is not ridiculously small and way higher than the Planck mass. An important property of the QNMs lies in the fact that the relevant value is the one of $M\omega$: it is the product of the mass by the frequency that has a given (complex) value. The  characteristics of the QNMs of a lighter BH are exactly the same than those of a heavier one, they are simply shifted to higher frequencies by the mass ratio. Some quantum corrections might explicitly break this scaling law. This is the case of the Hayward metric considered below.\\

It should first be pointed out that the work presented here aims at being quite generic and is not directly linked with the proposal  \cite{Haggard:2016ibp}. The plots previously shown can be used to get an estimate of the QNMs displacement for any model with a roughly gaussian modification to the metric. In addition, and very speculatively, it could be argued that the maximum possible time to be used to evaluate the mass (the higher the time, the higher the mass) is not necessarily bounded by the age of the Universe: in quantum gravity a ``bounce" is possible \cite{lqc9} and black holes could survive during this bounce \cite{Clifton:2017hvg}. In principe it is therefore conceivable that a time much larger than the inverse Hubble parameter could used \cite{Carr:2017wkz,Barrau:2017ukm}, leading to measurable quantum gravity effects in QNMs at much higher masses

\section{The Hayward metric}

Recently, an effective metric for Planck stars \cite{Haggard:2014rza} has been proposed in \cite{DeLorenzo:2014pta}. The idea is to cure two usual inconsistencies of most metrics: the absence of a correct treatment of the time dilatation between the center and infinity and the failure to reproduce 1-loop quantum corrections (as calculated {\it e.g.} in \cite{BjerrumBohr:2002kt}). As a step in this direction, the authors make use of the Hayward metric (revived in a quantum gravity context \cite{Frolov:2014jva}):
\begin{equation} 
\label{eq:F(r)M(r)}
F(r) = 1- \frac{2 m(r)}{r} \;.
\end{equation}
Several proposals were made for the function $m(r)$. We consider here the original version \cite{Hayward:2005gi} where 
\be\label{eq:M(r)}
m(r) = \frac{M\,r^3}{r^3 + 2 M L^2},
\ee
where $L$ has the dimensions of a length. We consider only the case where $L<\frac{4}{3\sqrt{3}}M$, otherwise there is no horizon. We have investigated the displacement of QNMs as a function of $L$ -- which intuitively quantifies the scale of ``quantumness" -- for a given mass. As expected, the minimal required value of $L$, for a given relative QNM move, is proportional to $M$. If we require $(\frac{\Delta\omega}{\omega})$ to be of the order of a few percent (that is in the ET sensitivity range), the minimum value of $L$ is of the order of 0.7 mass units. More specifically $(\frac{\Delta\omega}{\omega})\sim 5\%$ is achieved for $L/M\sim 0.72$. For a macroscopic BH, this is much larger than the Planck length and this means that in such approaches the quantum modifications would need to be extending substantially beyond the usually assumed length scale.

\section{Conclusion and prospects}

We have shown that if quantum gravity effects leak outside the horizon of a Schwarzschild black hole, the quasinormal modes can -- as expected -- be substantially modified. Using a gaussian truncation of the metric structure, we have studied the influence of all the parameters describing the perturbations. In particular, we have quantified the amplitude of the quantum bump required for observation. Using a toy-model, we have translated the derived values into an upper limit on the mass leading to observable effects.\\

In the future, this approach should be refined by considering a rotating black hole. It would also be important to estimate de possible degeneracies: could the change in frequency and damping rate mimic a usual BH of different mass and spin?\\

Finally, it would be welcome to consider more realistic metrics based on heuristic quantum gravity arguments, in particular based either on loop quantum gravity black holes (see \cite{Barrau:2018rts} for a review) of on string black holes (see \cite{Mathur:2008nj} for interesting new ideas).

\section{Acknowledgments}

K.M is supported by a grant from the CFM foundation. The authors thank deeply R.A. Konoplya who provided us with his code.

\bibliography{refs}

\begin{thebibliography}{44}
\expandafter\ifx\csname natexlab\endcsname\relax\def\natexlab#1{#1}\fi
\expandafter\ifx\csname bibnamefont\endcsname\relax
  \def\bibnamefont#1{#1}\fi
\expandafter\ifx\csname bibfnamefont\endcsname\relax
  \def\bibfnamefont#1{#1}\fi
\expandafter\ifx\csname citenamefont\endcsname\relax
  \def\citenamefont#1{#1}\fi
\expandafter\ifx\csname url\endcsname\relax
  \def\url#1{\texttt{#1}}\fi
\expandafter\ifx\csname urlprefix\endcsname\relax\def\urlprefix{URL }\fi
\providecommand{\bibinfo}[2]{#2}
\providecommand{\eprint}[2][]{\url{#2}}

\bibitem[{\citenamefont{Barrau}(2017)}]{Barrau:2017tcd}
\bibinfo{author}{\bibfnamefont{A.}~\bibnamefont{Barrau}},
  \bibinfo{journal}{Comptes Rendus Physique} \textbf{\bibinfo{volume}{18}},
  \bibinfo{pages}{189} (\bibinfo{year}{2017}), \eprint{1705.01597}.

\bibitem[{\citenamefont{Giddings and Psaltis}(2018)}]{Giddings:2016btb}
\bibinfo{author}{\bibfnamefont{S.~B.} \bibnamefont{Giddings}} \bibnamefont{and}
  \bibinfo{author}{\bibfnamefont{D.}~\bibnamefont{Psaltis}},
  \bibinfo{journal}{Phys. Rev.} \textbf{\bibinfo{volume}{D97}},
  \bibinfo{pages}{084035} (\bibinfo{year}{2018}), \eprint{1606.07814}.

\bibitem[{\citenamefont{Giddings}(2019)}]{Giddings:2019jwy}
\bibinfo{author}{\bibfnamefont{S.~B.} \bibnamefont{Giddings}}
  (\bibinfo{year}{2019}), \eprint{1904.05287}.

\bibitem[{\citenamefont{Akiyama et~al.}(2019{\natexlab{a}})}]{Akiyama:2019cqa}
\bibinfo{author}{\bibfnamefont{K.}~\bibnamefont{Akiyama}} \bibnamefont{et~al.}
  (\bibinfo{collaboration}{Event Horizon Telescope}),
  \bibinfo{journal}{Astrophys. J.} \textbf{\bibinfo{volume}{875}},
  \bibinfo{pages}{L1} (\bibinfo{year}{2019}{\natexlab{a}}).

\bibitem[{\citenamefont{Akiyama et~al.}(2019{\natexlab{b}})}]{Akiyama:2019eap}
\bibinfo{author}{\bibfnamefont{K.}~\bibnamefont{Akiyama}} \bibnamefont{et~al.}
  (\bibinfo{collaboration}{Event Horizon Telescope}),
  \bibinfo{journal}{Astrophys. J.} \textbf{\bibinfo{volume}{875}},
  \bibinfo{pages}{L6} (\bibinfo{year}{2019}{\natexlab{b}}).

\bibitem[{\citenamefont{Almheiri
  et~al.}(2013{\natexlab{a}})\citenamefont{Almheiri, Marolf, Polchinski, and
  Sully}}]{Almheiri:2012rt}
\bibinfo{author}{\bibfnamefont{A.}~\bibnamefont{Almheiri}},
  \bibinfo{author}{\bibfnamefont{D.}~\bibnamefont{Marolf}},
  \bibinfo{author}{\bibfnamefont{J.}~\bibnamefont{Polchinski}},
  \bibnamefont{and} \bibinfo{author}{\bibfnamefont{J.}~\bibnamefont{Sully}},
  \bibinfo{journal}{JHEP} \textbf{\bibinfo{volume}{02}}, \bibinfo{pages}{062}
  (\bibinfo{year}{2013}{\natexlab{a}}), \eprint{1207.3123}.

\bibitem[{\citenamefont{Almheiri
  et~al.}(2013{\natexlab{b}})\citenamefont{Almheiri, Marolf, Polchinski,
  Stanford, and Sully}}]{Almheiri:2013hfa}
\bibinfo{author}{\bibfnamefont{A.}~\bibnamefont{Almheiri}},
  \bibinfo{author}{\bibfnamefont{D.}~\bibnamefont{Marolf}},
  \bibinfo{author}{\bibfnamefont{J.}~\bibnamefont{Polchinski}},
  \bibinfo{author}{\bibfnamefont{D.}~\bibnamefont{Stanford}}, \bibnamefont{and}
  \bibinfo{author}{\bibfnamefont{J.}~\bibnamefont{Sully}},
  \bibinfo{journal}{JHEP} \textbf{\bibinfo{volume}{09}}, \bibinfo{pages}{018}
  (\bibinfo{year}{2013}{\natexlab{b}}), \eprint{1304.6483}.

\bibitem[{\citenamefont{Mazur and Mottola}(2004)}]{Mazur:2004fk}
\bibinfo{author}{\bibfnamefont{P.~O.} \bibnamefont{Mazur}} \bibnamefont{and}
  \bibinfo{author}{\bibfnamefont{E.}~\bibnamefont{Mottola}},
  \bibinfo{journal}{Proc. Nat. Acad. Sci.} \textbf{\bibinfo{volume}{101}},
  \bibinfo{pages}{9545} (\bibinfo{year}{2004}), \eprint{gr-qc/0407075}.

\bibitem[{\citenamefont{Mathur}(2008)}]{Mathur:2008nj}
\bibinfo{author}{\bibfnamefont{S.~D.} \bibnamefont{Mathur}}
  (\bibinfo{year}{2008}), \eprint{0810.4525}.

\bibitem[{\citenamefont{Giddings}(1992)}]{Giddings:1992hh}
\bibinfo{author}{\bibfnamefont{S.~B.} \bibnamefont{Giddings}},
  \bibinfo{journal}{Phys. Rev.} \textbf{\bibinfo{volume}{D46}},
  \bibinfo{pages}{1347} (\bibinfo{year}{1992}), \eprint{hep-th/9203059}.

\bibitem[{\citenamefont{Maldacena and Susskind}(2013)}]{Maldacena:2013xja}
\bibinfo{author}{\bibfnamefont{J.}~\bibnamefont{Maldacena}} \bibnamefont{and}
  \bibinfo{author}{\bibfnamefont{L.}~\bibnamefont{Susskind}},
  \bibinfo{journal}{Fortsch. Phys.} \textbf{\bibinfo{volume}{61}},
  \bibinfo{pages}{781} (\bibinfo{year}{2013}), \eprint{1306.0533}.

\bibitem[{\citenamefont{Barceló et~al.}(2017)\citenamefont{Barceló,
  Carballo-Rubio, and Garay}}]{Barcelo:2017lnx}
\bibinfo{author}{\bibfnamefont{C.}~\bibnamefont{Barceló}},
  \bibinfo{author}{\bibfnamefont{R.}~\bibnamefont{Carballo-Rubio}},
  \bibnamefont{and} \bibinfo{author}{\bibfnamefont{L.~J.} \bibnamefont{Garay}},
  \bibinfo{journal}{JHEP} \textbf{\bibinfo{volume}{05}}, \bibinfo{pages}{054}
  (\bibinfo{year}{2017}), \eprint{1701.09156}.

\bibitem[{\citenamefont{Haggard and Rovelli}(2015)}]{Haggard:2014rza}
\bibinfo{author}{\bibfnamefont{H.~M.} \bibnamefont{Haggard}} \bibnamefont{and}
  \bibinfo{author}{\bibfnamefont{C.}~\bibnamefont{Rovelli}},
  \bibinfo{journal}{Phys. Rev.} \textbf{\bibinfo{volume}{D92}},
  \bibinfo{pages}{104020} (\bibinfo{year}{2015}), \eprint{1407.0989}.

\bibitem[{\citenamefont{Haggard and Rovelli}(2016)}]{Haggard:2016ibp}
\bibinfo{author}{\bibfnamefont{H.~M.} \bibnamefont{Haggard}} \bibnamefont{and}
  \bibinfo{author}{\bibfnamefont{C.}~\bibnamefont{Rovelli}},
  \bibinfo{journal}{Int. J. Mod. Phys.} \textbf{\bibinfo{volume}{D25}},
  \bibinfo{pages}{1644021} (\bibinfo{year}{2016}), \eprint{1607.00364}.

\bibitem[{\citenamefont{Chirenti}(2018)}]{Chirenti:2017mwe}
\bibinfo{author}{\bibfnamefont{C.}~\bibnamefont{Chirenti}},
  \bibinfo{journal}{Braz. J. Phys.} \textbf{\bibinfo{volume}{48}},
  \bibinfo{pages}{102} (\bibinfo{year}{2018}), \eprint{1708.04476}.

\bibitem[{\citenamefont{Moulin et~al.}(2018)\citenamefont{Moulin, Martineau,
  Grain, and Barrau}}]{Moulin:2018uap}
\bibinfo{author}{\bibfnamefont{F.}~\bibnamefont{Moulin}},
  \bibinfo{author}{\bibfnamefont{K.}~\bibnamefont{Martineau}},
  \bibinfo{author}{\bibfnamefont{J.}~\bibnamefont{Grain}}, \bibnamefont{and}
  \bibinfo{author}{\bibfnamefont{A.}~\bibnamefont{Barrau}}
  (\bibinfo{year}{2018}), \eprint{1808.00207}.

\bibitem[{\citenamefont{Chandrasekhar}(1985)}]{Chandrasekhar:1985kt}
\bibinfo{author}{\bibfnamefont{S.}~\bibnamefont{Chandrasekhar}}, in
  \emph{\bibinfo{booktitle}{{Oxford, UK: Clarendon (1992) 646 p., OXFORD, UK:
  CLARENDON (1985) 646 P.}}} (\bibinfo{year}{1985}).

\bibitem[{\citenamefont{Moulin and Barrau}(2019)}]{Flora}
\bibinfo{author}{\bibfnamefont{F.}~\bibnamefont{Moulin}} \bibnamefont{and}
  \bibinfo{author}{\bibfnamefont{A.}~\bibnamefont{Barrau}}
  (\bibinfo{year}{2019}), \eprint{in prepration}.

\bibitem[{\citenamefont{Kokkotas and Schmidt}(1999)}]{Kokkotas:1999bd}
\bibinfo{author}{\bibfnamefont{K.~D.} \bibnamefont{Kokkotas}} \bibnamefont{and}
  \bibinfo{author}{\bibfnamefont{B.~G.} \bibnamefont{Schmidt}},
  \bibinfo{journal}{Living Rev. Rel.} \textbf{\bibinfo{volume}{2}},
  \bibinfo{pages}{2} (\bibinfo{year}{1999}), \eprint{gr-qc/9909058}.

\bibitem[{\citenamefont{Nollert}(1999)}]{Nollert:1999ji}
\bibinfo{author}{\bibfnamefont{H.-P.} \bibnamefont{Nollert}},
  \bibinfo{journal}{Class. Quant. Grav.} \textbf{\bibinfo{volume}{16}},
  \bibinfo{pages}{R159} (\bibinfo{year}{1999}).

\bibitem[{\citenamefont{Berti et~al.}(2004)\citenamefont{Berti, Cardoso, and
  Yoshida}}]{Berti:2004um}
\bibinfo{author}{\bibfnamefont{E.}~\bibnamefont{Berti}},
  \bibinfo{author}{\bibfnamefont{V.}~\bibnamefont{Cardoso}}, \bibnamefont{and}
  \bibinfo{author}{\bibfnamefont{S.}~\bibnamefont{Yoshida}},
  \bibinfo{journal}{Phys. Rev.} \textbf{\bibinfo{volume}{D69}},
  \bibinfo{pages}{124018} (\bibinfo{year}{2004}), \eprint{gr-qc/0401052}.

\bibitem[{\citenamefont{Dorband et~al.}(2006)\citenamefont{Dorband, Berti,
  Diener, Schnetter, and Tiglio}}]{Dorband:2006gg}
\bibinfo{author}{\bibfnamefont{E.~N.} \bibnamefont{Dorband}},
  \bibinfo{author}{\bibfnamefont{E.}~\bibnamefont{Berti}},
  \bibinfo{author}{\bibfnamefont{P.}~\bibnamefont{Diener}},
  \bibinfo{author}{\bibfnamefont{E.}~\bibnamefont{Schnetter}},
  \bibnamefont{and} \bibinfo{author}{\bibfnamefont{M.}~\bibnamefont{Tiglio}},
  \bibinfo{journal}{Phys. Rev.} \textbf{\bibinfo{volume}{D74}},
  \bibinfo{pages}{084028} (\bibinfo{year}{2006}), \eprint{gr-qc/0608091}.

\bibitem[{\citenamefont{Konoplya}(2003)}]{Konoplya:2003ii}
\bibinfo{author}{\bibfnamefont{R.~A.} \bibnamefont{Konoplya}},
  \bibinfo{journal}{Phys. Rev.} \textbf{\bibinfo{volume}{D68}},
  \bibinfo{pages}{024018} (\bibinfo{year}{2003}), \eprint{gr-qc/0303052}.

\bibitem[{\citenamefont{Schutz and Will}(1985)}]{Schutz:1985zz}
\bibinfo{author}{\bibfnamefont{B.~F.} \bibnamefont{Schutz}} \bibnamefont{and}
  \bibinfo{author}{\bibfnamefont{C.~M.} \bibnamefont{Will}},
  \bibinfo{journal}{Astrophys. J.} \textbf{\bibinfo{volume}{291}},
  \bibinfo{pages}{L33} (\bibinfo{year}{1985}).

\bibitem[{\citenamefont{Iyer and Will}(1987)}]{Iyer:1986np}
\bibinfo{author}{\bibfnamefont{S.}~\bibnamefont{Iyer}} \bibnamefont{and}
  \bibinfo{author}{\bibfnamefont{C.~M.} \bibnamefont{Will}},
  \bibinfo{journal}{Phys. Rev.} \textbf{\bibinfo{volume}{D35}},
  \bibinfo{pages}{3621} (\bibinfo{year}{1987}).

\bibitem[{\citenamefont{Iyer}(1987)}]{Iyer:1986nq}
\bibinfo{author}{\bibfnamefont{S.}~\bibnamefont{Iyer}}, \bibinfo{journal}{Phys.
  Rev.} \textbf{\bibinfo{volume}{D35}}, \bibinfo{pages}{3632}
  (\bibinfo{year}{1987}).

\bibitem[{\citenamefont{Kokkotas and Schutz}(1988)}]{Kokkotas:1988fm}
\bibinfo{author}{\bibfnamefont{K.~D.} \bibnamefont{Kokkotas}} \bibnamefont{and}
  \bibinfo{author}{\bibfnamefont{B.~F.} \bibnamefont{Schutz}},
  \bibinfo{journal}{Phys. Rev.} \textbf{\bibinfo{volume}{D37}},
  \bibinfo{pages}{3378} (\bibinfo{year}{1988}).

\bibitem[{\citenamefont{Konoplya et~al.}(2019)\citenamefont{Konoplya, Zhidenko,
  and Zinhailo}}]{Konoplya:2019hlu}
\bibinfo{author}{\bibfnamefont{R.~A.} \bibnamefont{Konoplya}},
  \bibinfo{author}{\bibfnamefont{A.}~\bibnamefont{Zhidenko}}, \bibnamefont{and}
  \bibinfo{author}{\bibfnamefont{A.~F.} \bibnamefont{Zinhailo}}
  (\bibinfo{year}{2019}), \eprint{1904.10333}.

\bibitem[{\citenamefont{Rovelli and Vidotto}(2014)}]{Rovelli:2014cta}
\bibinfo{author}{\bibfnamefont{C.}~\bibnamefont{Rovelli}} \bibnamefont{and}
  \bibinfo{author}{\bibfnamefont{F.}~\bibnamefont{Vidotto}},
  \bibinfo{journal}{Int. J. Mod. Phys.} \textbf{\bibinfo{volume}{D23}},
  \bibinfo{pages}{1442026} (\bibinfo{year}{2014}), \eprint{1401.6562}.

\bibitem[{\citenamefont{Barrau and Rovelli}(2014)}]{Barrau:2014hda}
\bibinfo{author}{\bibfnamefont{A.}~\bibnamefont{Barrau}} \bibnamefont{and}
  \bibinfo{author}{\bibfnamefont{C.}~\bibnamefont{Rovelli}},
  \bibinfo{journal}{Phys. Lett.} \textbf{\bibinfo{volume}{B739}},
  \bibinfo{pages}{405} (\bibinfo{year}{2014}), \eprint{1404.5821}.

\bibitem[{\citenamefont{Barrau et~al.}(2014)\citenamefont{Barrau, Rovelli, and
  Vidotto}}]{Barrau:2014yka}
\bibinfo{author}{\bibfnamefont{A.}~\bibnamefont{Barrau}},
  \bibinfo{author}{\bibfnamefont{C.}~\bibnamefont{Rovelli}}, \bibnamefont{and}
  \bibinfo{author}{\bibfnamefont{F.}~\bibnamefont{Vidotto}},
  \bibinfo{journal}{Phys. Rev.} \textbf{\bibinfo{volume}{D90}},
  \bibinfo{pages}{127503} (\bibinfo{year}{2014}), \eprint{1409.4031}.

\bibitem[{\citenamefont{Barrau et~al.}(2018{\natexlab{a}})\citenamefont{Barrau,
  Moulin, and Martineau}}]{Barrau:2018kyv}
\bibinfo{author}{\bibfnamefont{A.}~\bibnamefont{Barrau}},
  \bibinfo{author}{\bibfnamefont{F.}~\bibnamefont{Moulin}}, \bibnamefont{and}
  \bibinfo{author}{\bibfnamefont{K.}~\bibnamefont{Martineau}},
  \bibinfo{journal}{Phys. Rev.} \textbf{\bibinfo{volume}{D97}},
  \bibinfo{pages}{066019} (\bibinfo{year}{2018}{\natexlab{a}}),
  \eprint{1801.03841}.

\bibitem[{\citenamefont{Abbott et~al.}(2016)}]{Abbott:2016blz}
\bibinfo{author}{\bibfnamefont{B.~P.} \bibnamefont{Abbott}}
  \bibnamefont{et~al.} (\bibinfo{collaboration}{LIGO Scientific, Virgo}),
  \bibinfo{journal}{Phys. Rev. Lett.} \textbf{\bibinfo{volume}{116}},
  \bibinfo{pages}{061102} (\bibinfo{year}{2016}), \eprint{1602.03837}.

\bibitem[{\citenamefont{Abbott et~al.}(2018)}]{LIGOScientific:2018mvr}
\bibinfo{author}{\bibfnamefont{B.~P.} \bibnamefont{Abbott}}
  \bibnamefont{et~al.} (\bibinfo{collaboration}{LIGO Scientific, Virgo})
  (\bibinfo{year}{2018}), \eprint{1811.12907}.

\bibitem[{\citenamefont{Sathyaprakash et~al.}(2012)}]{Sathyaprakash:2012jk}
\bibinfo{author}{\bibfnamefont{B.}~\bibnamefont{Sathyaprakash}}
  \bibnamefont{et~al.}, \bibinfo{journal}{Class. Quant. Grav.}
  \textbf{\bibinfo{volume}{29}}, \bibinfo{pages}{124013}
  (\bibinfo{year}{2012}), \bibinfo{note}{[Erratum: Class. Quant.
  Grav.30,079501(2013)]}, \eprint{1206.0331}.

\bibitem[{\citenamefont{Ashtekar and Singh}(2011)}]{lqc9}
\bibinfo{author}{\bibfnamefont{A.}~\bibnamefont{Ashtekar}} \bibnamefont{and}
  \bibinfo{author}{\bibfnamefont{P.}~\bibnamefont{Singh}},
  \bibinfo{journal}{Class. Quant. Grav.} \textbf{\bibinfo{volume}{28}},
  \bibinfo{pages}{213001} (\bibinfo{year}{2011}), \eprint{1108.0893}.

\bibitem[{\citenamefont{Clifton et~al.}(2017)\citenamefont{Clifton, Carr, and
  Coley}}]{Clifton:2017hvg}
\bibinfo{author}{\bibfnamefont{T.}~\bibnamefont{Clifton}},
  \bibinfo{author}{\bibfnamefont{B.}~\bibnamefont{Carr}}, \bibnamefont{and}
  \bibinfo{author}{\bibfnamefont{A.}~\bibnamefont{Coley}},
  \bibinfo{journal}{Class. Quant. Grav.} \textbf{\bibinfo{volume}{34}},
  \bibinfo{pages}{135005} (\bibinfo{year}{2017}), \eprint{1701.05750}.

\bibitem[{\citenamefont{Carr et~al.}(2017)\citenamefont{Carr, Clifton, and
  Coley}}]{Carr:2017wkz}
\bibinfo{author}{\bibfnamefont{B.}~\bibnamefont{Carr}},
  \bibinfo{author}{\bibfnamefont{T.}~\bibnamefont{Clifton}}, \bibnamefont{and}
  \bibinfo{author}{\bibfnamefont{A.}~\bibnamefont{Coley}}
  (\bibinfo{year}{2017}), \eprint{1704.02919}.

\bibitem[{\citenamefont{Barrau et~al.}(2017)\citenamefont{Barrau, Martineau,
  and Moulin}}]{Barrau:2017ukm}
\bibinfo{author}{\bibfnamefont{A.}~\bibnamefont{Barrau}},
  \bibinfo{author}{\bibfnamefont{K.}~\bibnamefont{Martineau}},
  \bibnamefont{and} \bibinfo{author}{\bibfnamefont{F.}~\bibnamefont{Moulin}},
  \bibinfo{journal}{Phys. Rev.} \textbf{\bibinfo{volume}{D96}},
  \bibinfo{pages}{123520} (\bibinfo{year}{2017}), \eprint{1711.05301}.

\bibitem[{\citenamefont{De~Lorenzo et~al.}(2015)\citenamefont{De~Lorenzo,
  Pacilio, Rovelli, and Speziale}}]{DeLorenzo:2014pta}
\bibinfo{author}{\bibfnamefont{T.}~\bibnamefont{De~Lorenzo}},
  \bibinfo{author}{\bibfnamefont{C.}~\bibnamefont{Pacilio}},
  \bibinfo{author}{\bibfnamefont{C.}~\bibnamefont{Rovelli}}, \bibnamefont{and}
  \bibinfo{author}{\bibfnamefont{S.}~\bibnamefont{Speziale}},
  \bibinfo{journal}{Gen. Rel. Grav.} \textbf{\bibinfo{volume}{47}},
  \bibinfo{pages}{41} (\bibinfo{year}{2015}), \eprint{1412.6015}.

\bibitem[{\citenamefont{Bjerrum-Bohr et~al.}(2003)\citenamefont{Bjerrum-Bohr,
  Donoghue, and Holstein}}]{BjerrumBohr:2002kt}
\bibinfo{author}{\bibfnamefont{N.~E.~J.} \bibnamefont{Bjerrum-Bohr}},
  \bibinfo{author}{\bibfnamefont{J.~F.} \bibnamefont{Donoghue}},
  \bibnamefont{and} \bibinfo{author}{\bibfnamefont{B.~R.}
  \bibnamefont{Holstein}}, \bibinfo{journal}{Phys. Rev.}
  \textbf{\bibinfo{volume}{D67}}, \bibinfo{pages}{084033}
  (\bibinfo{year}{2003}), \bibinfo{note}{[Erratum: Phys.
  Rev.D71,069903(2005)]}, \eprint{hep-th/0211072}.

\bibitem[{\citenamefont{Frolov}(2014)}]{Frolov:2014jva}
\bibinfo{author}{\bibfnamefont{V.~P.} \bibnamefont{Frolov}},
  \bibinfo{journal}{JHEP} \textbf{\bibinfo{volume}{05}}, \bibinfo{pages}{049}
  (\bibinfo{year}{2014}), \eprint{1402.5446}.

\bibitem[{\citenamefont{Hayward}(2006)}]{Hayward:2005gi}
\bibinfo{author}{\bibfnamefont{S.~A.} \bibnamefont{Hayward}},
  \bibinfo{journal}{Phys. Rev. Lett.} \textbf{\bibinfo{volume}{96}},
  \bibinfo{pages}{031103} (\bibinfo{year}{2006}), \eprint{gr-qc/0506126}.

\bibitem[{\citenamefont{Barrau et~al.}(2018{\natexlab{b}})\citenamefont{Barrau,
  Martineau, and Moulin}}]{Barrau:2018rts}
\bibinfo{author}{\bibfnamefont{A.}~\bibnamefont{Barrau}},
  \bibinfo{author}{\bibfnamefont{K.}~\bibnamefont{Martineau}},
  \bibnamefont{and} \bibinfo{author}{\bibfnamefont{F.}~\bibnamefont{Moulin}},
  \bibinfo{journal}{Universe} \textbf{\bibinfo{volume}{4}},
  \bibinfo{pages}{102} (\bibinfo{year}{2018}{\natexlab{b}}),
  \eprint{1808.08857}.

\end{thebibliography}

 \end{document}